\documentclass[%
 aip,
 amsmath,amssymb,
preprint,%
]{revtex4-1}

\usepackage{graphicx}
\usepackage{amsmath}
\usepackage{xcolor}
\usepackage{etoolbox}

\DeclareRobustCommand*{\citen}[1]{%
  \begingroup
    \romannumeral-`\x 
    \setcitestyle{numbers}%
    \cite{#1}%
  \endgroup
}

\begin{document}

\title[]{Quantum bath augmented stochastic nonequilibrium atomistic simulations for molecular heat conduction}


\author{Renai Chen}%
\email{renaic@sas.upenn.edu}
\affiliation{Department of Chemistry, University of Pennsylvania, Philadelphia, PA  19104, USA}
\affiliation{Theoretical Division and Center for Nonlinear Studies, Los Alamos National Laboratory, Los Alamos, NM, 87544, USA}
\author{Mohammadhasan Dinpajooh}
\affiliation{Physical and Computational Sciences Directorate, Pacific Northwest National Laboratory, WA 99352 USA}
\author{Abraham Nitzan}
\affiliation{Department of Chemistry, University of Pennsylvania, Philadelphia, PA  19104, USA}
\affiliation{School of Chemistry, Tel Aviv University, Tel Aviv 69978, Israel}

\begin{abstract}
	Classical molecular dynamics (MD) has been shown
	to be effective in simulating heat conduction
	in certain molecular junctions since it inherently takes into account 
	some essential methodological components which are lacking 
	with quantum Landauer-type transport model, such as many-body full force-field interactions,
	anharmonicity effects and nonlinear responses for large temperature biases.
	However, the classical mechanics reaches its limit in the environments
	where the quantum effects are significant (e.g. with low-temperatures substrates,
	presence of extremely high frequency molecular modes).
	Here, we present an atomistic simulation methodology for molecular heat conduction
	that incorporates the quantum Bose-Einstein statistics into an ``effective temperature''
	in the form of modified Langevin equation.
	We show that the results from such a quasi-classical effective temperature (QCET) MD method deviates drastically 
	when the baths temperature approaches zero from classical MD simulations
	and the results converge to the classical ones when the bath approaches the high-temperature limit,
	which makes the method suitable for full temperature range.
	%
        In addition, we show that our quasi-classical thermal transport method can be used to model the conducting substrate layout 
	and molecular composition (e.g. anharmonicities, high-frequency modes).
	Anharmonic models are explicitly simulated via the Morse potential and compared to 
	pure harmonic interactions, to show the effects of anharmonicities under 
	quantum colored bath setups. Finally, the chain length dependence of heat conduction is examined for one-dimensional polymer chains placed in between quantum augmented baths.
\end{abstract}


\maketitle 

\section{Introduction}
Heat transport in molecular systems is of great importance
not only because it is an important subfield of nanoscale heat transfer
\cite{Cahill2003,Cahill2014,Benenti2017}
which promises optimized microscopic heat management,
but also for its significant roles in molecular electronics,
novel thermal-electric interactions and phononic computings.
\cite{cuevas2010molecular,Dubi2011,Segal2016arpc,Cui2017jcp,
Renai2017,Craven2018prl,Wang2007science,Rubtsova2015acr,
Pein2013jpca,Sadeghi2015nl,Li2012rmp}
Substrate-molecule-substrate type molecular junction (MJ) structures
are the most prevalent testing grounds for both theoretical and experimental studies of molecular heat conduction.
\cite{Segal2016arpc,Segal2003,Renai2020jpcl,Meier2014prl,Majumdar2017nl,Gaskins2015jpcc,Majumdar2015nl}
Recent experimental advances on single-molecule
junction (SMJ) high-precision heat conduction measurements,
\cite{Cui2019nature, Mosso2019nl}
has put an urgency of finding 
a comprehensive and reliable theoretical approach
and numerical implementation,
so that heat conductions in SMJ can be
accurately and systematically investigated. 

One of the dominant theoretical methodologies for thermal transport
in MJs is the Nonequilibrium Green's Function (NEGF) method.
In principle, such an approach can capture the full dynamics
of the transport properties at steady-state. 
This approach is straightforward to implement in harmonic models.
However, in practice, the harmonic approximation is often made
to lower the computational costs and simplify the calculations.
\cite{Segal2003,Cui2019nature,Kloeckner2016,
Cui2017science,Kloeckner2017prb_tuning,Kloeckner2018prb}
Though such an approximation may still be reasonable for low temperature environments, 
its applicability in higher temperatures or 
systems with high anharmonicities could be problematic.
On the other hand, the long-standing trajectory-based molecular dynamics (MD)
has been used to simulate thermal conductivities of various nanoscale materials.
\cite{Schelling2002prb,Sellan2010prb,
McGaughey2004ijhmt1,McGaughey2004ijhmt2,Chen2012jap,Sellan2010prb,Zhang2005jpcb,Tang2013apl,Dong2014scirep,Majumdar2017nl,Nguyen2010jcp,Hung2016jpcc}
Its rich pool of parameterized complex systems that include many-body interactions
and comparatively low computational costs
make it suitable for simulating realistic systems where anharmonic effects could be important.

Indeed, a GROMACS-based computational toolkit 
that utilizes stochastic baths and substrate filterings
has been recently developed
to simulate heat conduction across molecular junctions of different topologies
and has been shown to be accurate and reliable 
for room temperature settings.\cite{Sharony2020}
Nonetheless, the classical MD-based method loses its predicting power 
when the temperatures of the leads are pushed to a much lower end(e.g. 25K at one side of the junction)
and the conductance deviates greatly from the quantum calculations.\cite{Sharony2020,Renai2020jpcl}
One of the main reasons for such inaccuracy is the inability of 
classical mechanics in capturing 
the dominant quantum effect at low temperatures
which primarily arises from quantum Bose-Einstein distribution.

Recently, there have been research efforts aimed to introduce quantum bath effects
in non-equilibrium spin-boson (NESB) model for molecular heat transports,\cite{Carpio-Martinez2021jcp} 
in which significant differences are found between the samplings of classical and quantum baths.
In addition, methodological development efforts have been put into blending quantum effects
into classical MD using generalized Langevin equation\cite{Wang2007prl} (GLE)
and an attempt to implement such a method for the thermal transport processes
in molecular junctions has been reported\cite{Li2021prb}.
While such a quasi-classical approach seems to capture quantum effects
of heat conduction for certain molecular junctions,
the transformations between Fourier and time spaces
of the noises could lead to heavy computational overheads.
For large and realistic systems,
further methodological developments are needed.

In this paper, 
circumventing painstaking assumptions and approximations of using GLE,
we introduce a general mathematical framework that
starts from Wiener-Khinchin-Einstein theorem
to integrate quantum bath effects into MD simulations smoothly
in the thermal transport simulations in the molecules,
without sacrificing all the benefits 
inherited from classical simulations.
The quantum fluctuation-dissipation relation appears to partly resemble 
GLE, but the clear pathway of implementing the stochastic force
makes the method conceptually and practically easy to be implemented.
The implemented code contains the possibility to include colored noise. 
Such colored spectrum can be engineered based on the atomic parameters 
and coupling strengths depending on specific molecular systems under study.
We will first elaborate on the detailed formalism of
incorporating such effects through an effective temperature reconstruction
to the Langevin dynamcis.
Then the method is demonstrated using toy models
by looking into monatomic relaxation and diatomic heat conduction.
In the meantime, a colored bath is simulated explicitly
with tunable parameters of the bath internal degrees of freedom 
(e.g coupling constants).
We finally test the quantum plus colored bath model
on heat conduction with harmonic and anharmonic diatomic molecules
and polymeric chain molecules,
laying out foundations for the upcoming implementations
of this general method into MD packages.

\section{Methods and formalism}
\subsection{\label{sec:method}Formalism for quasi-classical effective temperature molecular dynamics simulation}
We start out with the Langevin's equation,
for a single (bath) particle at position $x$
(assuming one dimensional)
under the potential $V(x)$.
\begin{equation}
	m\ddot{x}(t) = -\frac{\partial V(x)}{\partial x} - m\gamma\dot{x} + R(t).
\end{equation}
$\gamma$ is the friction term,
and $R(t)$ is the random fluctuation force acting on the particle.
The friction $\gamma$ and the random noise $R$ are related by the fluctuation-dissipation theorem 
\begin{equation}
	\label{eqn:FD}
	\langle R(t)R(t^\prime)\rangle = 2m\gamma k_B\mathcal{T}\delta(t-t^\prime),
\end{equation}
where $k_B$ is the Boltzmann constant
and $\mathcal{T}$ refers to the surrounding thermal environment or bath temperature.

Consider the random fluctuation term in more detail.
Suppose we observe the system in time interval
$0\le t\le T$ and expand $R(t)$ in Fourier series.
\begin{equation}
	R(t) = \sum_{n=-\infty}^{n=\infty}R_ne^{i\omega_nt}
\end{equation}
\begin{equation}
	\omega_n = \frac{2\pi n}{T},\quad n=0,\pm1,\pm2\cdots
\end{equation}
\begin{equation}
	R_n=\frac{1}{T}\int_{0}^{T}dt R(t)e^{-i\omega_nt}
\end{equation}
The power spectrum can be written as\cite{Nitzan2006chemical}:
\begin{equation}
	I_R(\omega)=\lim_{T\rightarrow\infty}\left(\frac{\sum_{n\in W_{\Delta\omega}}\langle|R_n|^2\rangle}{\Delta\omega} \right);
\end{equation}
where
\begin{equation}
	W_{\Delta\omega}=\{ n|\omega-\Delta\omega/2\le 2\pi n/T\le \omega+\Delta\omega/2\},
\end{equation}
which can further change to
\begin{equation}
	I_R(\omega)=\lim_{T\rightarrow\infty}\sum_{n\in W_{\Delta\omega}}\langle|R_n|^2\rangle\delta(\omega-\omega_n);
	\quad \omega=\frac{2\pi n}{T}.
\end{equation}
$I_R(\omega)\Delta\omega$ is the intensity of the random noise
in the frequency range $\omega\cdots\omega+\Delta\omega$,
which is obtained by summing up the magnitudes of different components $R_n^2$.

The Wiener-Khintchine theorem states:
\begin{equation}
	I_R(\omega) = \frac{1}{2\pi}\int_{-\infty}^{\infty}dt e^{-i\omega t}\langle R(t)R(0)\rangle.
\end{equation}
Together with Eqn.(\ref{eqn:FD}),
we get
\begin{equation}
	\label{}
	I_R=\frac{m\gamma k_B\mathcal{T}}{\pi}
\end{equation}
Here we want to find an approximate coarse-grained frequency resolution of the heat flux.
To this end we divide the relevant frequency range $\{0\cdots \omega_{max}\}$ into $N$ segments $\Delta\omega=\omega_{max}/N$.
For each such segment the random noise can be taken as,
\begin{align}
\label{eq:Rn_Fourier}
	R_n(t)=&\sqrt{\Delta\omega\frac{m\gamma k_B\mathcal{T}}{\pi}}\left(e^{i(\omega^{(n)}t+\phi_n)}+e^{-i(\omega^{(n)}t+\phi_n)}\right)\\ \nonumber
	&=2\sqrt{\Delta\omega\frac{m\gamma k_B\mathcal{T}}{\pi}}\cos(\omega^{(n)}t+\phi_n),
\end{align}
where $\omega^{(n)}$ is best chosen in the middle of the segment, that is $\omega^{(n)}=(n-1/2)\Delta\omega$ and $\phi_n$ is the random phase,
which will be averaged over multiple trajectories
of different values.

Alternatively, each segment can be further divided into more than one frequency, say $\{\omega^{(n)}_j\},j=1\cdots M_n$ evenly distributed in the segments.
In this case, the driving corresponding to the frequency segment n
(between $\omega_{n-1}$ and $\omega_n$) is
\begin{equation}
	R_n(t)
	=2\sqrt{\frac{\Delta\omega}{M_n}\frac{m\gamma k_B\mathcal{T}}{\pi}}\sum_j\cos(\omega^{(n)}_jt+\phi_j).
 \label{eqn:Rn}
\end{equation}
The approximation lies in the assumption 
that the currents obtained from these drivings are additive,
so that if we act together with two segments n and n+1 
the heat current will be the sum of the individual contributions. 
This is certainly so in harmonic systems but is only an approximation for anharmonic ones. 
The reason why the approximation might be feasible 
is that contributions from different segments 
will come with random phases 
so that mixed signal might average to zero 
when average over phases is done.

\begin{figure*}
\centering
\includegraphics[width=1\textwidth]{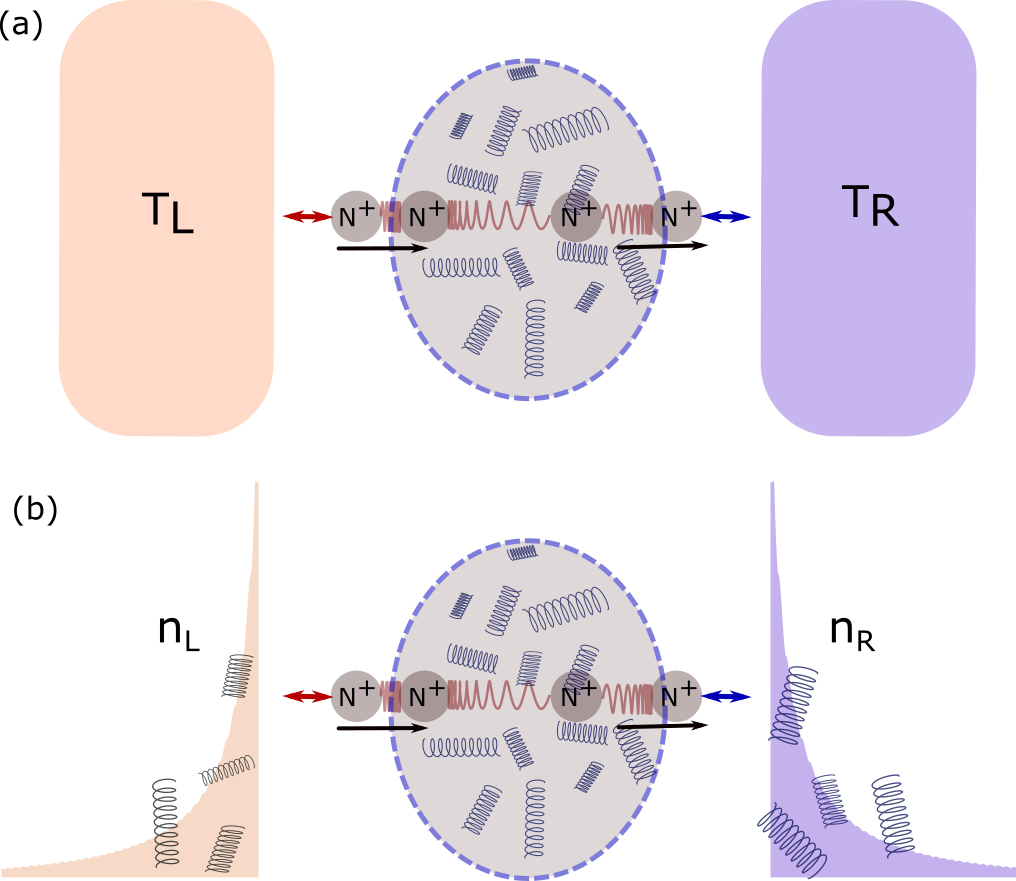}
\caption{Schematic diagrams of heat transport in molecular junctions across temperature biases.
	(a) When the thermal baths are classical, meaning they do not differentiate different bath modes of different frequencies;
	(b) How the actual quantum baths look like: different bath modes weigh differently with respect to the corresponding temperatures,
	distributed according to Bose-Einstein phonon population distributions.}
\label{fig:schematic_bath}
\end{figure*}

Qualitatively, one might conceptualize such an effective temperature
represents frequency weighed thermal contribute from single bath mode
versus uniform constant temperatures for classical MD (See Fig. \ref{fig:schematic_bath} (a)).
Together with 
the interplay between individual bath modes
and molecular modes,
the modified random force embodies the collective effect of all the vibrational modes
within the thermal baths (like shown in Fig. \ref{fig:schematic_bath} (b))
and its effect on the overall heat conduction,
in a way analogous to Bose-Einstein population in the Landauer's transport formalism.~\cite{Segal2016arpc}
Taking more frequencies is just like taking smaller segments.
Therefore, the final expression for the random noise can be written as

\begin{align}
	\label{eqn:Rt}
	R_n
	=2\sqrt{\Delta\omega\frac{m\gamma k_B\mathcal{T}_{eff}(\omega^{(n)})}{\pi}}\cos(\omega^{(n)}t+\phi_n),\\ \nonumber
	R(t) = \sum_nR_n(t),
\end{align}
where
\begin{equation}
	\label{eqn:Teff}
	\mathcal{T}_{eff}(\omega)=\frac{\hbar\omega/k_B}{e^{\hbar\omega/k_B\mathcal{T}}-1}.
\end{equation}
Different realizations of the random noise 
come from different choices of $\phi_n$ for different choices of random noise.
Note that now we have reformulated the random noise in our quasi-classical approach,
we can do standard heat conduction without calculating frequency-resolved transmission coefficients, and the details about the heat current calculations (atomic local heat fluxes) can be found in a previous paper.\cite{Sharony2020} All that is required is to replace the random noise with a similar one 
but with a frequency-dependent effective temperature noting that the coupling of the system to the quantum bath involves explicitly the classical bath friction ($\gamma$), the range of bath frequencies ($\omega^{(n)}$), and Bose-Einstein distribution for the effective bath temperature ($\mathcal{T}_{eff}$) all of which capture the bath quantum effects. In what follows, this method is called quasi-classical effective temperature (QCET) method.
The QCET approach is shown to be accurate and computationally effective for simple models in molecular junctions and has been also demonstrated for one-dimensional polymeric systems.

\subsection{\label{sec:relaxation} Atomistic design of colored bath}

We will also consider heat transport between thermal baths characterized by the frequency dependence and cutoff given by the Debye model
and their spectral properties. 
In such a scenario,
the quantum augmented bath model with modes of different frequencies contribute differently
will provide more insights into the thermal transport dynamics than plain white-noise based classical MD approaches.
Specifically, the spectrum behaves as $\omega^2$ for small $\omega$'s,
and has a cutoff frequency ($\omega_D$) when $\omega\rightarrow\infty$.
One way to characterize such a colored bath spectrum
is through mathematical filtering,\cite{Nitzan1978jcp}
which implicitly incorporates the color of the bath
into the Generalized Langevin equation.
Another category of methods is explicitly engineering bath atomic properties and compositions (e.g. masses, bond strengths),
which works well for toy model studies\cite{Kalantar2021pre}
and is much more straightforward computationally.
For the purpose of this study,
we will choose the latter strategy as laid out specifically below.

\begin{figure*}
	\centering
	\includegraphics[width=1\textwidth]{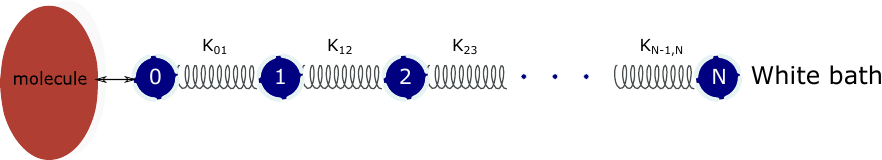}
	\caption{
	Schematic drawing of the Debye bath composed
	of N atoms, 
	attaching a bridge molecule on the left
	and white bath on the right.
	}
	\label{fig:molecule_bath}
\end{figure*}

Consider the case where the colored bath consists of N atoms arranged in a linear chain (See Fig.~\ref{fig:molecule_bath}).
We want the atom number 0 to represent a Debye bath effect.
We start by regarding the system without the molecule.
The potential for bilinear interactions can be written as 
(in 1-dimensional notations):

\begin{equation}
	V(x_0,x_1,\cdots,x_N) = \frac{1}{2}\sum_{j=1}^N\sum_{l=1}^NK_{jl}x_jx_l
\end{equation}

\begin{align}
	\frac{d^2x_n}{dt^2}=-\frac{1}{M_n}[K_{n,n+1}x_{n+1}+K_{n,n-1}x_{n-1}+K_{nn}x_n];\\ \nonumber
 n\neq 0,N
\end{align}
The last atom (N) is connected to white noise,
\begin{align}
	\frac{d^2x_N}{dt^2}=&\frac{1}{M_N}[K_{N,N-1}x_{N-1}+K_{NN}x_N]\\ \nonumber
    &-\gamma\frac{dx_N}{dt}+\frac{1}{M_N}R(t), \\ \nonumber
	&\langle R(t)R(0)\rangle=2M_N\gamma k_BT\delta(t).
\end{align}

After mathematical transformations and derivations (see details in Appendix~\ref{app:Debye_formalism}),
the spectrum density of the explicit bath,
\begin{equation}
	g(\omega) = |[\mathbf{M}^{-1}]_{0N}|^2\omega^2\frac{3\gamma}{\pi^2},
\end{equation}
where $\mathbf{M}$ is the dynamical force and frequency related matrix defined in Appendix ~\ref{app:Debye_formalism}.

With parameters of force constants (i.e. $k_{ij}$) and the coupling ($\gamma$),
we can fit the density of modes into Debye-shape (with a normalization factor C),
\begin{equation}
	g(\omega) = C\frac{\omega^2}{1+(\omega/\omega_D)^{2(N+1)}},
	\label{eqn:g_omega_fit}
\end{equation}
which shows in the same form of the math filter method.\cite{Nitzan1978jcp}

As an example of the above formalism,
we will use
two layers (one atom for each layer) of representative
bath atoms that give rise to the eighth power of $\omega$
when fitted into Eq. \ref{eqn:g_omega_fit}
(with specific tunable parameters),
which could be a good estimator
for the purpose of simple model investigations.
It can be shown the specific values for the tunning parameters can be calculated (details in Appendix~\ref{app:Debye_formalism}).

\begin{table}[tbh]
\centering
\caption{Molecular dynamics simulation parameters for various models studied in this work in the reduced units. Using the symbol $\bar{\omega}$ for the diatomic molecule's internal vibrational frequency, the frequency mode associated with the reduced Debye frequency is set as $\omega_{\mathrm{D}}= 4\bar{\omega}$ (value calculated in Eq. \ref{eq:Debye_cutoff_define}). The force constants inside the Debye baths, $k_1$ and $k_2$ are the reduced spring constants associated with the bath atoms arranged in a linear chain, the last of which is connected to the white bath with a reduced friction coefficient of $\gamma$. See Appendix \ref{app:Debye_formalism} for detailed derivations.}
\begin{tabular}{ccccc}
\hline
    Quasiclassical Bath & $k_1$ & $k_2$ & $\gamma$ & Random Noise  \\
     \hline
  & 17.21$\bar{\omega}^2$ & 14.95$\bar{\omega}^2$ & 9.26$\bar{\omega}$  & Eq. \ref{eqn:Rt}\\
       \hline
    System: Diatomic Molecule & $k_0$  &  &  & \\
     \hline
  & $\bar{\omega}^2$  & &  & \\
     \hline
    System: 1D-Polymer & $k_0$ (Eq. \ref{polymer_eqn}) &  &  & \\
     \hline
  & 11.37$\bar{\omega}^2$  & &  & \\
       \hline
    System-Bath: Harmonic & $k_{R}$ & $k_{L}$ &  & \\
     \hline
  &  $0.23\bar{\omega}^2$ & $0.23\bar{\omega}^2$ &  & \\
     \hline
    System-Bath: Morse & $\alpha$ & $D$ & $2\alpha^2D$ & \\
     \hline
  &  $0.33\sqrt{\bar{\omega}/\hbar}$ & $1.06\hbar\bar{\omega}$ &$0.23\bar{\omega}^2$  & \\
       \hline
\end{tabular}
\label{pot_param}
\end{table}


\clearpage
\section{Simple Applications: Relaxation of a single harmonic oscillator}
\begin{figure*}[tbh]
	\centering
	\includegraphics[width=1\textwidth]{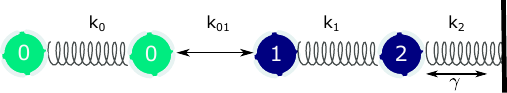}
	\caption{
	Schematic drawing of the diatomic molecule 
	attached to a colored bath represented by two harmonically bonded atoms and a thermal reservoir.
	The molecule is represented on the left by the number 0, 
	connected to the bath atoms,
	which are parameterized according to Table~\ref{pot_param}.
	}
	\label{fig:diatomic_relaxation_model}
\end{figure*}

We start by applying the QCET approach 
to the relaxation dynamics of a single oscillator, representing a diatomic molecule, coupled to a harmonic heat bath as shown in Fig. \ref{fig:diatomic_relaxation_model}.

Making use of the analytical expressions for 
both classical and quantum equilibrium energies 
of harmonic and anharmonic (Morse) oscillators (see Fig. S2 and S3 in  SI. Note that SI is separate from the appendices),
we simulate the relaxation processes of both
the classical MD approach 
and the QCET 
dynamics approach (Eq. \ref{eqn:Rt}) we have developed and 
compare them to the available analytical results.

The simulation of QCET,
which considers quantum effect of the reservoirs,
can account for both high and low temperature limits 
at equilibrium,
by encoding an effective temperature that incorporates
quantum boson distribution into the molecular dynamics (see Fig. S1 in SI).
To show the effect of colored-Debye bath and anharmonicity
within the capacity of the QCET method, 
we simulate the energy relaxation
of a diatomic molecule coupled to the bath surface.
Here, we use different interactions
and parameters (see Table \ref{pot_param} and Fig.~\ref{fig:diatomic_relaxation_model})
to show the rates of energy relaxation for such a model
with respect to bath temperature changes.

\begin{figure}
	\centering
	\includegraphics[width=1\columnwidth]{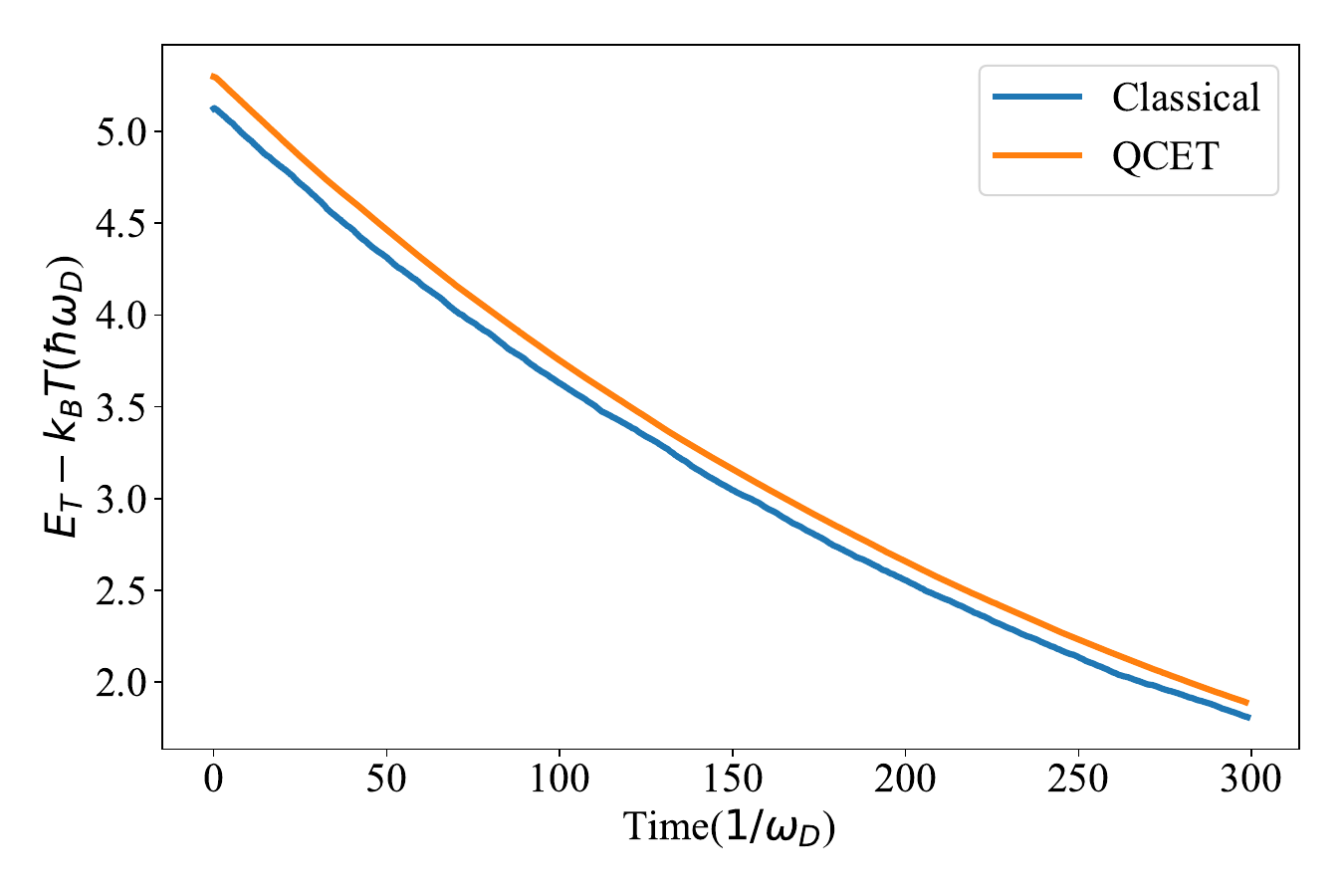}
	\caption{
	Examples of energy relaxation rates of the diatomic molecule shown 
	in Fig.~\ref{fig:diatomic_relaxation_model}, 
	with classical and QCET method.
	The internal vibration frequency is 1/4 of Debye cutoff 
	and temperature is 0.2 ($\hbar\omega_D/k_B$).
	The horizontal axis is the evolution time.
	The slopes of the logarithmic fits of the energy changes (vertical axis values) are
	0.000209 for the classical 
	and 0.000207 for the quantum,
	which correspond to their relaxation rates 
	respectively.
	}
	\label{fig:rate_plot_examples}
\end{figure}

Two types of interactions are taken
between the molecule and the bath 
(i.e. between atom 0 and 1).
a). Harmonic ($k_{01}=0.23\bar{\omega}$ as default);
b). Anharmonic (Morse potential details in Section I and Section III in SI).
Figure~\ref{fig:rate_plot_examples} shows energy relaxation
for the harmonic oscillator within the range of Debye spectrum.
Initial states of the oscillator
are sampled from the same total energy
with random displacements and velocities.
The relaxation energy is calculated as 
the difference between the statistical ensemble averages
of the time-dependent total energy changes 
and the energy when the system 
is equilibrated to the bath.
While in the classical case the temperature is constant,
the effective temperature for the QCET 
comes from the collective effect of phonons in the quantum bath that follows Bose-Einstein distribution.
That is why the quasiclassical
appears slightly higher than the classical line,
because the equilibrium energy
of the augmented quantum bath is smaller
caused by overall lower effective temperature.
As the energy decays exponentially,
we can do logarithmic fits to 
get the relaxation rates from the slopes.
It is shown that the classical and quantum cases
align with each other well
($\sim$ 0.0002 in Fig.~\ref{fig:rate_plot_examples}).
However, it is not exactly the same for anharmonic interface potentials
(see Section IV in SI for detailed results).
At higher vibrational frequencies, the anharmonic (Morse) potential generates higher rates than the harmonic potential when connected to a Debye-colored bath, shown from the QCET simulations. 

\clearpage


\section{Heat conduction between Debye thermal baths}

\begin{figure}[tbh]
	\centering
	\includegraphics[width=0.8\columnwidth]{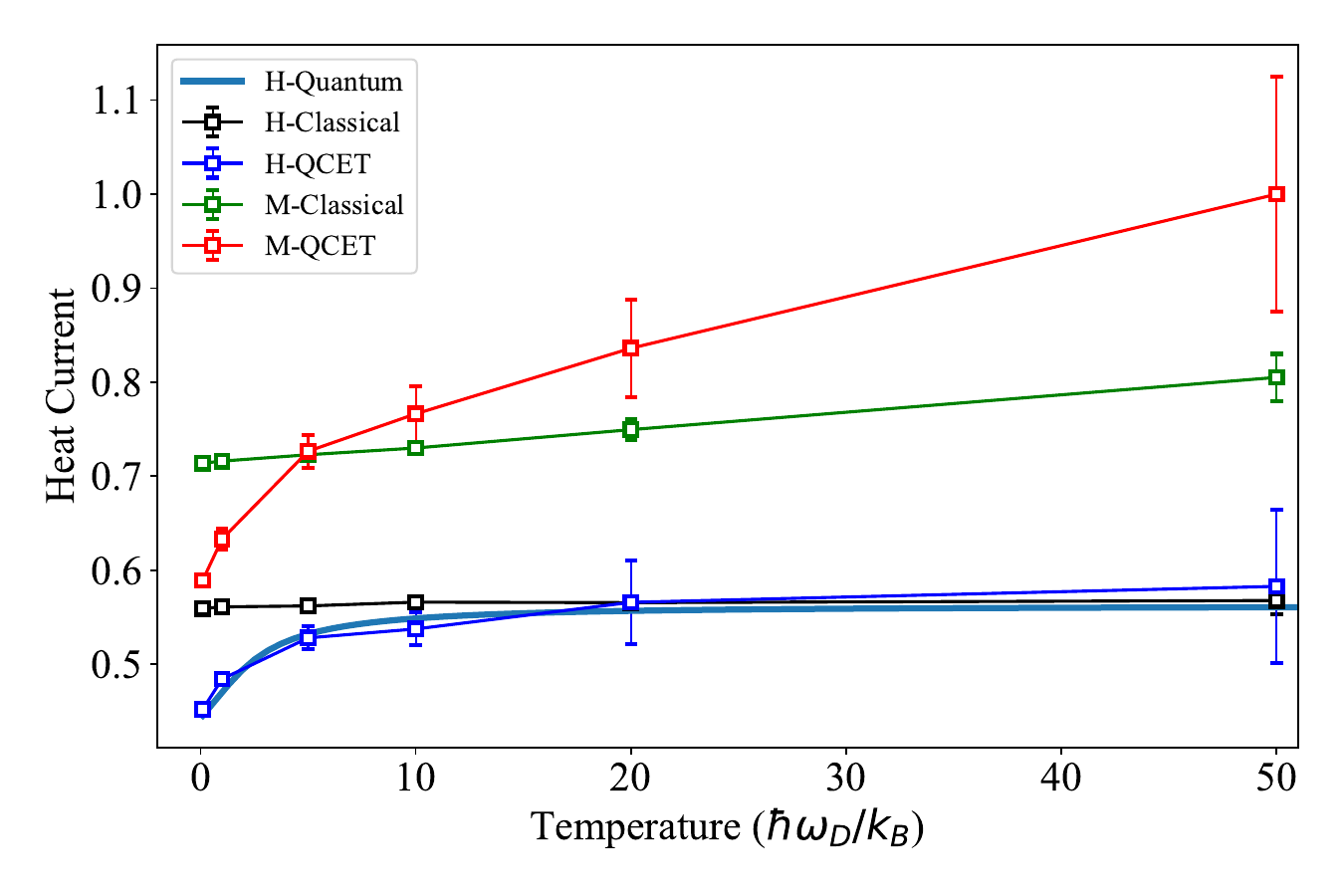}
	\caption{
	Heat currents of the diatomic molecular system
	connected to engineered Debye baths (specified in Fig. \ref{fig:diatomic_relaxation_model})
	one on the left and one on the right,
	with two different interactions between the system and baths,
	namely, harmonic (represented as H in the legend) and Morse (one example of anharmonic system-bath interactions, represented as M in the legend, 
	see Section III in SI for details).
 Quantum denotes Landauer-baseed quantum calculations.
	The parameters for left and right baths are the same (Table \ref{pot_param}),
	except the temperatures (the left bath temperatures are shown on the x-axis,
	and the right bath is always 20 $\hbar\omega/k_B$ higher than the left bath).
	The internal vibrational frequency for the diatomic molecule is 
	1/4 of the Debye cutoff (Eq. \ref{eq:Debye_cutoff_define}) of the baths,
	and the coupling between the bath and the molecule is $k_{01}=0.007\omega^2_\mathrm{D}$.
	For the Morse (anharmonic) interaction, the parameters are chosen such that $k_{01}=2\alpha^2D$.
	(see Section III in SI).
	The y-axis is normalized by the current of anharmonic quantum data at 70 $\hbar\omega/k_B$.
	The error bars represent Standard Error.\cite{SE}
	}
	\label{fig:heat_current_Debye}
\end{figure}

In this Section, the heat conduction is investigated for a diatomic  molecular system and a one-dimensional polymeric system all under engineered Debye-colored bahts (see Section \ref{sec:relaxation}).
Figure~\ref{fig:heat_current_Debye} shows the results
from classical MD simulations and QCET simulations for a diatomic molecular system.
All the parameters are set to be the same for Debye bath as described in Section~\ref{sec:relaxation},
only now we have two baths with different temperatures.

A few observations can be made for the diatomic molecule heat conduction with harmonic couplings to the baths in Fig. \ref{fig:heat_current_Debye} (lower three lines):
(a) The heat current from the classical MD approach
is almost T-independent across the low, middle to high 
bath temperatures, as long as the biases are kept the same.
(b) The Landauer's calculation reduces to about 
half of the value at lower temperature limit
compared to the constant temperature case
(the reason it does not go to zero
is that thermal bias always provides transporting phonons
in the right bath even when the left bath
has close-to-zero temperature),
and catches up gradually as temperature rises.
The trend of the QCET
aligns well with the Landauer's results.
(c) All three approaches converge to the same values,
as temperature of the left bath becomes high,
which is expected that quantum effect reduces
to its minimum at the classical (high-T) limit.

The Landauer's calculations depict accurate behaviors
for pure harmonic systems for the full temperature range,
while the classical MD results act
as its approximation for high-T limit
(similar behaviors can be shown 
with white baths as well for diatomic small molecules. See Fig. S4 in SI for details).
By showing the agreement of the 
QCET method
to the Landauer's calculation,
we confirm 
indeed that this method is capable of capturing full dynamics
of the molecular heat conduction,
through carefully engineering the bath properties
to introduce the Bose-Einstein distribution
of quantum mechanical phonons.

Nonetheless, for heat conduction in anharmonic systems, one will need to rely soley on simulational approaches for accurate depictions of heat transport features.
After replacing harmonic molecule-bath couplings with anharmonic (Morse) couplings (detailed Hamiltonains are shown in Section VI in SI),
the currents (top two lines) increase from 30\% to 60\% for all bath temperatures,
indicating the higher-order anharmonicities participate the thermal transport processes
and enhance the overall conducting effects. 
More interestingly, 
comparing between the classical anharmonic and QCET anharmonic results,
there is an increase for high temperature conduction for the quantum data,
which could be a combined effects from quantum statistical distributions
and anharmonic interactions from outside the Debye spectrum.
Such an enhancement which is absent from the classical counterparts
also agrees with the observations from the energy relaxation processes (see Fig. S2 and S3 in SI),
which further infers under the QCET model,
we are able to capture anharmonic system-bath coupling effects and quantum populations of molecular heat transport 
that are not available to either full quantum harmonic calculations (e.g. Landauer)
or classical MD simulations.
It is worth noting that Ref. \citen{Li2021prb} attributes the slight decrease 
of higher-temperature conduction for alkanes to anharmonicity in the molecules,
while in fact for their junction systems, with multiple layers of substrates
such deduction could also come from inelastic scattering events at the interface, which increase the resistance of the junctions to heat flow.
In a relevant study, Xufei {\it et al.}\cite{Xufei-2014} report the increase in the thermal conductance of monoatomic or diatomic lattices as the anharmonicity is increased. However, they report that tuning the anharmonicity right at the interface does not show obvious effects on the the thermal conductance. 
Therefore, there are too many contributing factors interweaving with each other
that it is not possible to explain the outcome from a dominant factor.
We show here through our simple anharmonic model, 
anharmonicities do not necessarily contribute to the reduced
thermal conduction at higher temperatures.
On the contrary, Morse-type potentials result in 
enhancement of conduction,
which could be illustrated by our newly developed QCET MD simulations.

\clearpage

Finally, the heat conduction is investigated for one-dimensional polymer chains placed between two colored baths (See Section I of SI for the Hamiltonian of this type of junction).
The one-dimensional polymer chains consists of $N$ units with the same mass ($m$) and spring constant ($k_0$) leading to the following equations of motion for the $n$th unit in the polymer chain:

\begin{equation}
m \ddot x_n =
\begin{cases}
 -2k_0 x_1 + k_0 x_{2} + F_L , \;\;\; n=1 \\
 k_0 x_{n-1} - 2k_0 x_n + k_0 x_{n+1} , \;\;\; 1<n<N \\
 -2k_0 x_N + k_0 x_{N-1} + F_R , \;\;\; n=N \\
\end{cases}
\label{polymer_eqn}
\end{equation}
where $F_L$ and $F_R$ represent the left and right bath forces to the closest units/atoms in the one-dimensional polymer, respectively. The same equations of motion for the (quantum augmented) bath as presented in Section \ref{sec:method} is used with two atoms to represent the bath (see Fig. \ref{fig:diatomic_relaxation_model}) and the polymer-bath coupling parameter as well as the temperature difference between baths are the same as the ones used for the diatomic case above (see Table \ref{pot_param}). To address the significance of the quantum augmented colored bath, one may compare the heat currents to the ones obtained in a classical colored bath, which is implemented by reducing the effective temperature in Eq. \ref{eqn:Teff} to the bath temperature. 

\begin{figure}[tbh]
	\centering
	\includegraphics[width=1\columnwidth]{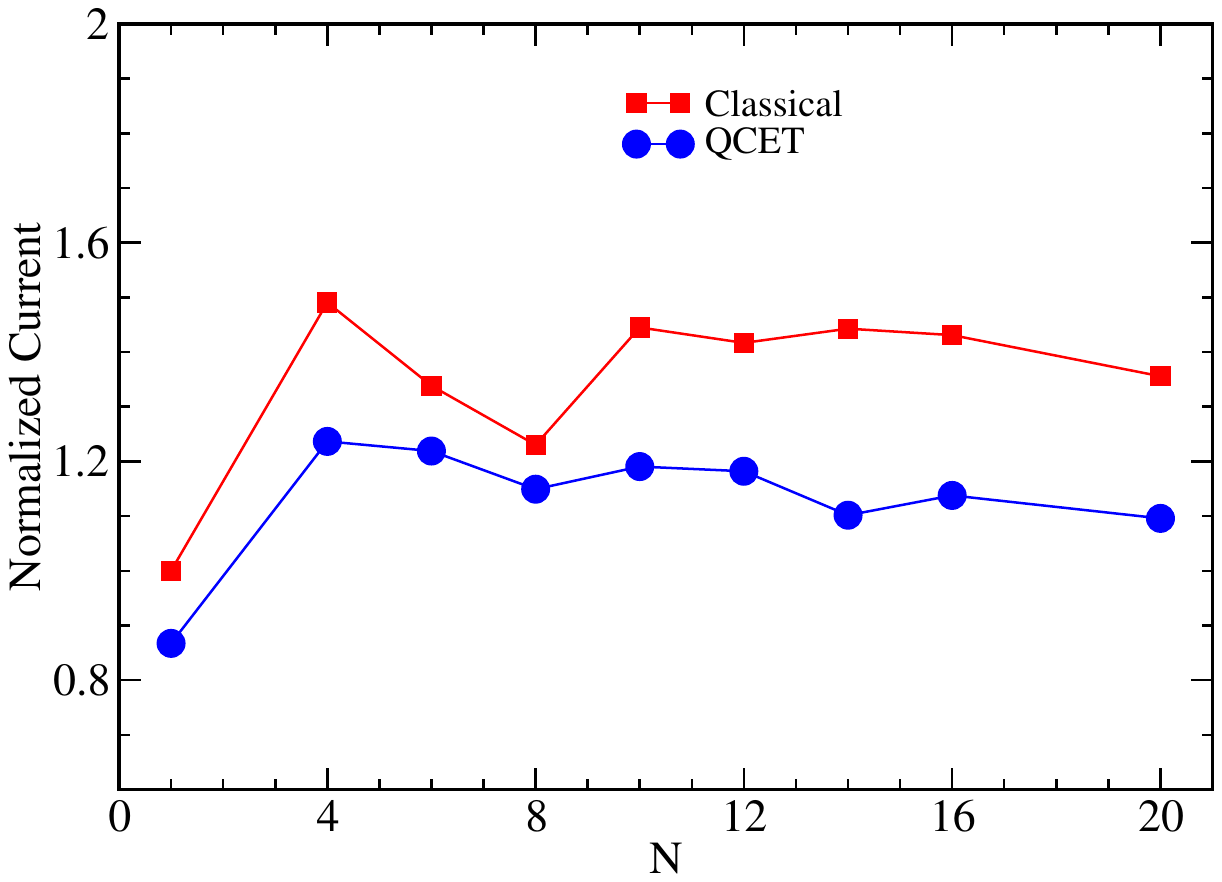}
	\caption{Normalized heat current at a reduced temperature of 20 $\hbar\omega/k_B$ for the lower bath 
(with bias being 2 $\hbar\omega/k_B$ ) for a one-dimensional polymer chain with various chain lengths coupled to classical or quasiclassical baths (see Fig. \ref{fig:diatomic_relaxation_model}). All the currents are normalized with respect to the heat current value of a molecule with $N=1$ coupled to the classical bath.}
	\label{fig:Ndep}
\end{figure}

The blue filled circles in Fig. \ref{fig:Ndep} show the dependence of the heat current on chain length for one-dimensional polymer chains consisting of $1–20$ units coupled to quantum augmented colored bath at a temperature of 20 $\hbar\omega/k_B$ for the lower bath 
(with bias being 2 $\hbar\omega/k_B$ ). 
As can be seen, the heat current has minimal length dependent for $N>10$, while for short
chains, a rise of the heat conductance with chain length is observed. 
The red filled squares in Fig. \ref{fig:Ndep} show the chain length dependence for one-dimensional polymer chains coupled to classical colored baths. All the results are normalized with the heat current value for the one-dimensional polymer chain with $N=1$ coupled with the classical colored bath. 
As can be seen, a similar trend is observed for the chain length dependence of polymer chains coupled to the classical bath noting that the heat currents from the classical baths are the upper heat current values when compared to the ones obtained from the quantum bath,
which shows the intrinsic working of Bose-Eninstein quantum populations in lowering the overall effective temperatures.

\clearpage
\section{Conclusion}
As we have pointed out in the development of stochastic nonequilibrium simulation
of molecular heat conduction~\cite{Sharony2020},
the inadequacy of describing thermal population of phononic modes
may have been the main reason leading to the failure of 
accurate depiction of low-temperature thermal conductance of single-molecule junctions
under classical MD.
Indeed, in this article
we have augmented the conventional atomistic simulation for molecular heat conduction
with quantum phononic baths,
in which the quantum thermal transport features 
are effectively incorporated into the dynamics
while at the same time the advantages  
of classical MD (e.g. full force-field interactions) is retained.
Such a successful augmentation has been first demonstrated through monatomic equilibration and diatomic molecular junction heat conduction.
In both scenarios we have seen 
the asymptotic behaviors of the quantum bath method
at the high-temperature limit always converge
to the classical approaches,
while at the low-temperature end 
they align with quantum analytical expressions
and Landauer's transport calculations.
We have thus overcome the shortcomings of classical MD
in simulating the molecular heat conduction when quantum effects may be significant,
with relatively small added computational complexity.
We have also introduced colored bath into the quantum bath method
to account for spectral features of contacting substrates (e.g. gold)
in the junctions and full-range of vibrational modes in the molecular bridges.
Our results in Fig. \ref{fig:heat_current_Debye} indicate the anharmonicity in the systems (modeled by Morse potential)
might increase or decrease the heat conductance
because of the interactions and scatterings of all the normal
modes in the systems,
which could be utilized to modulate heat conduction
by introducing or hybridizing different molecular structures.
The chain length dependence of heat current is also investigated for one-dimensional polymer chains coupled to quantum and classical colored baths. It is found that the heat currents for the colored classical baths are the upper values for the quantum augmented baths and both show minimal chain length dependence for polymer chains with more than $10$ units while the chain length dependence for polymer chains with less than $10$ units is significant.
For the future work direction,
the implementation of this method to the existing MD simulation package 
will be done for the full exploration
of molecular thermal transport covering all extreme conditions
(e.g. heat conduction of large systems like polymer chains\cite{Hadi2022jcp} at low and high ambient temperatures).
Such efforts may
verify and guide experimental studies in these directions
and prepare for the future applications in molecular phononics.

\clearpage

\section{SUPPLEMENTARY INFORMATION}
The the Hamiltonian of the junctions as well as the complementary analytical expressions for one-dimensional oscillators, anharmonic effects on the relaxation of Morse oscillators, diatomic molecule white-noise heat conduction, and Debye bath features are presented in the supplementary information.

\section{ACKNOWLEDGMENTS}
This work was supported by the U.S. National Science Foundation under Grant No. CHE1953701 and by the University of Pennsylvania.
R.C. acknowledges the support from the Center for Nonlinear Studies (CNLS) as a postdoctoral fellow at the Los Alamos National Laboratory (LANL). 
M. D. acknowledges the support from the U.S. Department of Energy, Office of Basic Energy Sciences, Division of Chemical Science, Geosciences, and Biosciences at Pacific Northwest National Laboratory (PNNL).

\section{Data Availability}
The data that supports the findings of this study are available within the article and its supplementary material. Source code implementations are also available from the corresponding author upon reasonable request.

\appendix


\section{\label{app:dimensionless}Dimensionless units}
All the dimensionless variables are defined as follows
in terms of harmonic oscillator frequency (included Morse potential parameters):
\begin{align}
\begin{split}
	\label{eqn:dimensionless}
	\bar{x}=\frac{x}{\sqrt{\frac{\hbar}{m\omega}}} \\
	\bar{v}=\frac{v}{\sqrt{\frac{\hbar\omega}{m}}} \\
	\bar{\omega}_n = \frac{\omega_n}{\omega}\\
	\bar{t} = \omega t \\
	\bar{\gamma} = \frac{\gamma}{\omega} \\
	\bar{T}=\frac{\hbar\omega}{k_BT} \\
	\bar{T}^{(n)}_{eff} = \frac{\bar{\omega}_n}{e^{\bar{\omega}_n/\bar{T}-1}}\\
	\bar{D} = \frac{D}{\hbar\omega}\\
	\bar{\alpha}=\alpha\sqrt{\frac{\hbar}{m\omega}}
\end{split}
\end{align}

\section{\label{app:Debye_formalism}Derivations of parameters for explicit layers Debye bath model}

In mass-weighted coordinates, one gets
\begin{align}
	\frac{d^2y_n}{dt^2}=-(k_{n,n+1}y_{n+1} +k_{n,n-1}y_{n-1}+k_{nn}y_n);\\ \nonumber
 \quad n\neq 0,N \\ 
	\frac{d^2y_0}{dt^2}=-(k_{0,1}y_1+k_{00}y_0)\\
	\frac{d^2y_N}{dt^2}=-(k_{N,N-1}y_{N-1}+k_{NN}y_N)-\gamma\frac{dy_N}{dt}+\rho(t),
\end{align}
where we have defined,
\begin{align}
	y_n=\sqrt{M_n}x_n,\quad k_{ij}=\frac{K_{ij}}{\sqrt{M_iM_j}},\\ \nonumber k_{nn}=\frac{K_{nn}}{M_n},\quad \rho(t)=\frac{R(t)}{\sqrt{M_N}},\\
	\langle \rho(t)\rho(0)\rangle=2\gamma k_BT\delta(t).
\end{align}
Now as we transform to Fourier space,
\begin{equation}
	f(t) = \tilde{f}(\omega)e^{-i\omega t}+\tilde{f}(-\omega)e^{i\omega t}
\end{equation}

We have,
\begin{align}
	\omega^2\tilde{y}_0 &= k_{01}\tilde{y}_1+k_{00}\tilde{y}_0 \\
	\omega^2\tilde{y}_n &= k_{n,n+1}\tilde{y}_{n+1}+k_{n,n-1}\tilde{y}_{n-1}+k_{nn}\tilde{y}_n \\
	\omega^2\tilde{y}_N &= k_{N,N-1}\tilde{y}_{N-1}+k_{NN}\tilde{y}_N-i\omega\gamma\tilde{y}_N-\tilde{\rho}(\omega).
\end{align}
We may further put them into matrix form.
\begin{equation}
	\mathbf{My}=\mathbf{r},
\end{equation}
where
\begin{equation}
	\mathbf{r}=\begin{bmatrix}
		0\\
		0\\
		\vdots\\
		\tilde{\rho}
	\end{bmatrix}
\end{equation}

\begin{equation}
\mathbf{M} = \begin{bmatrix}
	k_{00}-\omega^2 & k_{01} & \dots & 0\\
	k_{10} & k_{11}-\omega^2 & k_{12}  & 0\\
    \vdots & \vdots &\ddots & \vdots\\
    0 &  \dots &   k_{N,N-1}  & k_{NN}-\omega^2-i\omega\gamma
    \end{bmatrix}
\end{equation}
The matrix can be inverted analytically,
\begin{equation}
	\tilde{y}_0(\omega)=[\mathbf{M}^{-1}]_{0N}\tilde{\rho}(\omega)
\end{equation}
whence we have
\begin{equation}
	\langle|\tilde{\dot{y}}_0(\omega)|^2\rangle=|[\mathbf{M}^{-1}]_{0N}|^2\omega^2\langle|\tilde{\rho}(\omega)|^2\rangle
\end{equation}
$\tilde{\rho}(\omega)$ is the Fourier transform of the random noise $\rho$ and its squared averaging equal to \cite{Nitzan2006chemical}
\begin{equation}
	\langle|\tilde{\rho}(\omega)|^2\rangle=\frac{\gamma k_BT}{\pi},
\end{equation}
note the mass is unity in the mass-weighted representation.
We then have
\begin{equation}
	\langle|\tilde{\dot{y}}_0(\omega)|^2\rangle=|[\mathbf{M}^{-1}]_{0N}|^2\omega^2\frac{\gamma k_BT}{\pi}
\end{equation}
The LHS is the Fourier transform \cite{Nitzan2006chemical} of the velocity correlation function of the atom in the bath seen by the molecule.

When we connect the bath spectrum density to the v-v correlation~\cite{Nitzan2006chemical}
\begin{equation}
	g(\omega) = \frac{3mN}{\pi k_BT}\int_{-\infty}^{\infty}dt\langle \dot{x}(0)\dot{x}(t)\rangle e^{-i\omega t}.
\end{equation}
We have the expression that takes into account the force constants matrix.
\begin{equation}
	g(\omega) = |[\mathbf{M}^{-1}]_{0N}|^2\omega^2\frac{3\gamma}{\pi^2}
\end{equation}

For two-atoms bath (see Fig.~\ref{fig:diatomic_relaxation_model})
the left atom of the bath was taken to couple to the molecule:

\begin{equation}
\mathbf{M}=
\begin{bmatrix}
k_{00} - \omega^2 & -k_1 \\
-k_1 & k_{11}-i\gamma\omega-\omega^2 
\end{bmatrix}
\end{equation}
\begin{equation}
k_{00} = k_1,\:
k_{11} = k_1+k_2
\end{equation}

\begin{equation}
	det(\mathbf{M})=\omega^4+i\gamma\omega^3-(2k_1+k_2)\omega^2-i\gamma k_1\omega+k_1k_2
\end{equation}

\begin{equation}
	\mathbf{M}^{-1}=\frac{1}{det(\mathbf{M})}
\begin{bmatrix}
k_1+k_2-i\gamma\omega-\omega^2 & k_1 \\
k_1 & k_1 - \omega^2
\end{bmatrix}
\end{equation}

\begin{equation}
	|\mathbf{M}^{-1}|_{12}=\frac{k_1}{det(\mathbf{M})}
\end{equation}

The denominator of $(|\mathbf{M}^{-1}|_{12})^2$ is
\begin{equation}
	\omega^8+c_6\omega^6+c_4\omega^4+c_2\omega^2+c_0
\end{equation}
where
\begin{align}
	c_6&=\gamma^2-2(2k_1+k_2) \\
	c_4&=(2k_1+k_2)^2+2k_1k_2-2\gamma^2k_1 \\
	c_2&=\gamma^2k^2_1-2k_1k_2(2k_1+k_2)  \\
	c_0&=k^2_1k^2_2
\end{align}
Under the condition, $c_6=c_4=c_2=0$ and $k_1\neq0$ and $k_2\neq0$, we fit the parameters 
\begin{equation}
	k_1 = 0.151388,\quad
	k_2 = 0.131483,\quad
	\gamma=0.868517
	\label{eqn:two_Debye_parameter_2}
\end{equation}
We can then go back to the expression of the density of modes,
\begin{equation}
	g(\omega)=(|\mathbf{M}^{-1}|_{12})^2\omega^2\frac{3\gamma}{\pi^2}
\end{equation}
and plot $g(\omega)$.
If we assume 
\begin{equation}
	g(\omega)=C\frac{\omega^2}{1+(\frac{\omega}{\omega_D})^8},
\end{equation}
then we have
\begin{equation}
\label{eq:Debye_cutoff_define}
	\omega_D=(c_0)^{1/8}\approx0.375126
\end{equation}
The spectrum is tested and confirmed with its Fourier transform 
of the auto-correlation functions 
to reflect the Debye cut-off characteristics.





\clearpage

\bibliography{j,references}

\end{document}